\documentclass{aa}
\usepackage{graphicx}
\begin{document}
\title{A seismological analysis of $\delta$ Scuti stars in the Pleiades cluster}

\author{L. Fox Machado \inst{1} \and
       F. P\'erez Hern\'andez \inst{1} \fnmsep \inst{2} \and 
       J.C. Su\'arez \inst{3}\fnmsep
               \thanks{also associate researcher at Observatoire de Paris.} \and  
       E. Michel \inst{4} \and
       Y. Lebreton \inst{5}
}

\offprints{L. Fox, \email{lfox@iac.es}}

\institute{Instituto de Astrof\'{\i}sica de Canarias, E-38205 La
Laguna, Tenerife, Spain
 \and Departamento de Astrof\'{\i}sica, Universidad de La Laguna, Tenerife,
Spain
 \and Instituto de Astrof\'{\i}sica de Andaluc\'{\i}a (CSIC), E-180080 Granada,
Spain
  \and Observatoire de Paris, LESIA, UMR 8109, F-92195 Meudon,
France
\and Observatoire de Paris, GEPI, UMR 8111, F-92195 Meudon,
France}

\date{Received 8 July 2005 / Accepted 12 September 2005}

\abstract{A comparison between the oscillation frequencies of six
multi-periodic $\delta$ Scuti stars of the Pleiades cluster and
the eigenfrequencies of rotating stellar models that match the
corresponding stellar parameters has been carried out. The
assumption that all the stars considered have some common
parameters, such as metallicity, distance or age, is imposed as a
constraint. As a result, we have a best fit solution associated
with a cluster metallicity of [Fe/H]$\simeq 0.067$, an age between
$70 \times 10^{6}$ and $100 \times 10^{6}\,$yr and a distance
modulus of $m_{V}-M_{V}=5.60$--5.70 mag. All the stars were found
to oscillate mainly in non-radial, low degree, low order $p$
modes. Estimates of mass and rotation rates for each star are
also obtained. }

\maketitle


\section{Introduction \label{sec:intro}}

The study of $\delta$ Scuti stars is expected to provide an
important diagnostic tool for the structure and evolution of
intermediate-mass stars at the main sequence evolution
and thick-shell hydrogen burning stages. Having masses between 1.5 and
2.5 $M_{\odot}$, these pulsating variables develop convective
cores during their central hydrogen-burning phases that make them
suitable for investigating the hydrodynamic processes occurring in
the core.

Although most of the general properties of these stars are well
understood, the application of seismological methods to derive their
properties has proved to be a difficult task. The problems
concerning the modelling of $\delta$ Scuti stars have been
discussed in details in recent reviews (e.g.\ Christensen-Dalsgaard
\cite{christensen}; Kjeldsen \cite{kjeldsen}). The most acute seem
to be the mode selection mechanism and the influence of rotation.
Concerning the former, although there is fairly good agreement
between the observed frequency range and that derived from
instability calculations (e.g.\ Michel et al.\ \cite{michel1}), far
more modes than observed are predicted to be unstable. 
While it is expected that the forthcoming
asteroseismology space mission COROT (Baglin et al.\ \cite{baglin})
will be able to  disentangle the whole spectra of $\delta$ Scuti
stars, for the moment the
mechanism that could cause an amplitude limitation in some
modes is still unknown. As a consequence, mode identification
based on the comparison between theoretical and observed
frequencies is difficult to attain in general. On the other hand,
although several observational techniques for mode identification
have been developed in recent years (Watson \cite{watson};
Garrido et al.\ \cite{garrido}; Aerts \cite{aerts}; Kenelly \&
Walker \cite{kenelly}; Viskum et al.\ \cite{viskum}), in only a few
cases have these techniques  been successfully applied.

It is well known that most $\delta$ Scuti stars
are rapid rotators. This has important effects on the
mode frequencies, both directly as a consequence of the changes that must be
introduced in the oscillation equations and indirectly through the changes in the
equilibrium models.
Therefore, successful analysis of  rotating $\delta$ Scuti stars
requires that rotation should be taken  into account consistently  at
all levels in the analysis.

Despite these problems, in recent years several 
attempts have been made to interpret the observed spectra
of $\delta$ Scuti stars (Goupil et al.\ \cite{goupil}; P\'erez
Hern\'andez et al.\ \cite{perez1}; Guzik \& Bradley \cite{guzik};
Pamyatnykh et al.\ \cite{pamyatnykh}; Bradley \& Guzik
\cite{bradley}; Hern\'andez et al.\ \cite{hernandez}; Su\'arez et
al.\ \cite{suarez1}). Here we address the problem of mode
identification for stars in an open cluster. These stars are
expected to have a common age and chemical composition.
Furthermore, the distance to the cluster is usually known with
high accuracy. The constraints imposed by the cluster parameters
have proved to be very useful when modelling an ensemble of
$\delta$ Scuti stars. The best examples of such studies are found
for the variables in the Praesepe cluster (Michel et al.\
\cite{michel1}; Hern\'andez et al.\ \cite{hernandez}). In a similar
way, we consider here several $\delta$ Scuti stars of the Pleiades
cluster and search for a best fit solution in the sense of a set
of stellar parameters that allows the simultaneous modelling of
all the stars considered, and that satisfies all the observables,
including the frequencies.

The paper is organized as follows. In Sect.~\ref{sec:stars}, we
describe the main characteristics of the Pleiades cluster and the
six $\delta$ Scuti stars under study. The modelling, the range of
parameters and the calculations of the eigenfrequencies are
discussed in Sect.~\ref{sec:models}. The details of the comparison
between observed and theoretical frequencies are discussed in
Sect.~\ref{sec:comp}. The results and their
discussion are given in Sect.~\ref{sec:results}. Finally, we
present our conclusions in Sect.~\ref{sec:conclusions}.

\section{The observational material \label{sec:stars}}

The Pleiades (M45) is a young ($\sim 75$--100\, Myr), Population I
cluster. Because of its proximity, the observational parameters of the
Pleiades have been intensively studied. In particular, the
metallicity of the cluster is estimated to be between $-0.14 \leq$
[Fe/H] $\leq +0.13$ (e.g. $-$0.034 $\pm$ 0.024 Boesgaard \& Friel
\cite{boesgaard}, 0.0260 $\pm$ 0.103 Cayrel de Strobel
\cite{cayrel}, $-$0.11 $\pm$ 0.03 Grenon \cite{grenon}),

Until recently, there was a dispute regarding the distance of the
Pleiades cluster from the Earth. While the determination of the
cluster distance  from direct parallax measurements of
HIPPARCOS gives 116.3$\pm^{3.3}_{3.2}$~pc
($m_{V}-M_{V}= 5.33\,\pm\, 0.06$ mag, Mermilliod et al.\
\cite{mermilliod}), the previously accepted distance, which is
based on comparing the cluster's main sequence with that of nearby
stars, was  $\approx$130 pc corresponding to a distance modulus of
about 5.60 mag (e.g.\ 5.65 $\pm$ 0.38 mag [Eggen \cite{eggen}],
5.60 $\pm$ 0.05 mag [Pinsonneault et al.\ \cite{pinsonneault}],
5.61 $\pm$ 0.03 mag [Stello \& Nissen \cite{stello}]).
As  will be discussed in Sect.~\ref{sec:results}, the distance derived
in the  research presented here agrees with the latter value.

\begin{table*}[ht]
  \caption{Observational properties of the target stars}
  \begin{tabular}{lccccccc}
  \hline
  Star        & HD &  ST  & $m_{V}$&$B-V$ & $E(B-V)$ & $V \sin\, i$ & $\beta$ \\
              &    &      &        &      &       & $(\mathrm{km\, s}^{-1})$&  \\
  \hline
  -       & 23628 & A7V   &  $7.66\pm0.03 $& $+0.211\pm0.005$ &$0.063 \pm0.008$ & $ 150 \pm15$& 2.884  \\
  V650 Tau& 23643 & A3V   &  $7.77\pm0.02 $& $+0.157\pm0.006$& $0.027\pm0.008$ &  $175\pm18$ & 2.862  \\
  -       & 23194 & A5V   &  $8.06\pm0.02 $& $+0.202\pm0.010$& $0.076\pm0.008$ & $20 \pm2$& 2.881  \\
  V624 Tau& 23156 & A7V   &  $8.23\pm0.01$ & $+0.250\pm0.005$& $0.046\pm0.008$ & $ 20 \pm2$& 2.823  \\
  V647 Tau& 23607 & A7V   &  $8.26\pm0.01$ & $+0.255 \pm0.001$&$0.057 \pm0.008$ & $ 20 \pm2$& 2.841  \\
  V534 Tau& 23567 & A9V   &  $8.28\pm0.02$ & $+0.360\pm0.001$ &$ 0.084\pm0.008$ & $ 90\pm 9$& 2.788  \\

  \hline
  \end{tabular}
  \label{tab:stars}
\end{table*}

To date, six $\delta$ Scuti stars are known in
the Pleiades. In a survey of variability in the
cluster, Breger (\cite{breger}) found   $\delta$ Scuti type oscillations
in four stars. The remaining two were recently reported to be
$\delta$ Scuti stars  by Koen (\cite{koen1}) and Li et al.\ (\cite{li}).
Some observational properties of these stars are given in
Table~\ref{tab:stars}.  The projected rotational
velocities, $v\sin  i$, were obtained from Morse et al.\
(\cite{morse}) and Uesugi \& Fukuda (\cite{uesugi}). A 10\%
uncertainty in these quantities has been assumed. The spectral
types (ST) and $\beta$ parameters were obtained from the SIMBAD database in
Strasbourg (France). The errors in the photometric parameters $(B-V)$
and $m_{V}$ are estimated from the dispersion between different
measurements of these quantities. The errors in reddening are
taken from Breger (\cite{breger1}). We shall use these
uncertainties in Section 3.2.

In recent years, the $\delta$ Scuti stars of the Pleiades cluster have been
observed in several campaigns of the STEPHI multi-site network
(Michel et al.\ \cite{michel2}). The information on
the oscillation frequencies of the stars used here has
been obtained from those campaigns and are published elsewhere
(Belmonte \& Michel \cite{belmonte1}; Michel et al.\
\cite{michel4}; Liu et al.\ \cite{liu}; Li et al.\ \cite{li};
Li et al.\ \cite{li1}; Fox-Machado et al.\ \cite{fox}).
The frequency peaks detected
with a confidence level above 99\% are summarized
in Table~\ref{tab:freq}. We note that the frequency resolution in a typical
STEPHI campaign (three weeks) is $\Delta \nu \sim 0.5\, \mu$Hz.

\begin{table}[!t]
\begin{center}
\caption{Observed frequencies used in this work. The data were obtained from
different STEPHI campaigns as detailed in the text.}
\begin{tabular}{lc|lc}
\hline \hline
star   & $\nu$   & star& $\nu $     \\
& $(\mu {\rm Hz})$&  &   $(\mu {\rm Hz})$ \\
\hline
HD 23628& 191.8 & V624 Tau &242.9    \\
    & 201.7  &&409.0    \\
    & 376.6  &&413.5   \\\cline{1-2}
V650 Tau&197.2&&416.4    \\
    &292.7&& 451.7   \\
    &333.1&& 489.4   \\
    &377.8 &&529.1   \\\hline
HD 23194&533.6&V647 Tau&236.6 \\
&574.9&&304.7    \\\cline{1-2}
V534 Tau &234.2 &&315.6    \\
    &252.9&&374.4    \\
    &307.6&&405.8     \\
    &377.9 &&444.1     \\
    &379.0 &&\\
    &448.1 &&  \\
    &525.0&&  \\\hline
\end{tabular}
\end{center}
\label{tab:freq}
\end{table}

\begin{figure}[!t]
\resizebox{\hsize}{!} {\includegraphics{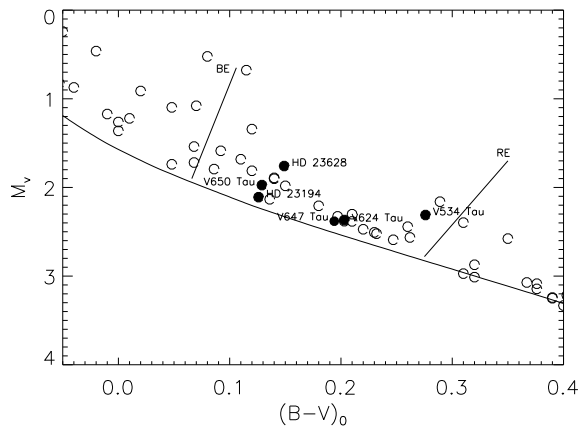}}
\caption{HR diagram of the Pleiades cluster. Only the region around the instability
strip is shown. The target stars are represented with filled circles.
The blue and red edges of the instability strip are indicated by continuous lines
and were taken from Rodr\'{\i}guez et al.\
(\cite{rodriguez}). An isochrone of 70 Myr is also shown.}
\label{fig:dscutis}
\end{figure}

Figure~\ref{fig:dscutis} illustrates a colour--magnitude diagram of the Pleiades
cluster in the region where the target stars are located.
The filled circles corresponds to the $\delta$ Scuti stars.
The apparent magnitudes, $V$, colours, $(B-V)$ and
membership used are contained in the WEBDA database (Mermilliod
\cite{mermilliod1}). The Pleiades shows
differential reddening with significant excess in the southwest with an
average value of $E(B-V) \approx 0.04$ (e.g.\
van Leeuwen \cite{vanleeuwen}; Breger \cite{breger1}; Hansen-Ruiz
\& van Leeuwen \cite{hansen}; Pinsonneault et al.\
\cite{pinsonneault}). In particular, we adopted the reddening for
individual stars as given by Breger (\cite{breger1}) if present, otherwise
the average of $E(B-V) = 0.04$ was used. Stars known to be spectroscopic binaries
were rejected.

Given the young age of the Pleiades cluster, we expect  all the
$\delta$-Scuti stars to be at an early evolutionary stage on the main sequence.
From the figure it follows that
while V647 Tau and V624 Tau have similar masses,
V650 Tau, HD 23194 and HD 23628 are slightly more massive.
V534 Tau, located near the red edge of the instability
strip, is the coolest star in the ensemble.

\section{The modelling \label{sec:models}}

In this section we shall explain how we have computed the stellar models,
the corresponding eigenfrequencies and how we have constrained the stellar
parameters used in the forward comparison with the observed frequencies.

\subsection{The stellar models}

We have considered stellar models
with input physics appropriate to the mass range covered by
$\delta$ Scuti stars. In particular, the stellar models were computed
using the CESAM evolutionary code (Morel \cite{morel}).
The nuclear reaction rates are from Caughlan \& Fowler
(\cite{caughlan}), the equation of state is from Eggleton et al.\
(\cite{eggleton}), the opacities from the OPAL project (Iglesias
\& Roger \cite{iglesias}) complemented at low temperatures by
Kurucz data (\cite{kurucz}), and  the
atmosphere is computed using the Hopf's $T(\tau)$ law (Mihalas
1978). The convection is described according
to the classical mixing-length theory (MLT hereafter). In particular,
we have considered an MLT parameter of $\alpha_{\rm MLT} =
l/H_{p}=1.52$, where $H_{p}$ is the pressure scale-height.
We have considered a fixed value for  $\alpha_{\rm  MLT}$ because this parameter is not expected
to have any significant influence on the global structure of the evolution models
considered for our target stars.

We have considered different sets of initial chemical  compositions
in order to cover the Pleiades [Fe/H] observed range (see
Section~\ref{sec:stars}) and different initial He abundances. The
values used are given in Table~\ref{tab:metal}. For each initial
chemical composition we have computed models with and without core
overshooting,  $\alpha_{{\rm ov}}=0.20$ in the former case 
(Schaller et al.\ \cite{schaller}).

\begin{table}[!t]
\begin{center}
\caption{Metallicity of the models. [Fe/H] = $\log (Z/X)_{\ast}- \log (Z/X)_{\odot}$,
with $(Z/X)_{\odot}=0.0245$ from Grevesse \& Noels (\cite{grevesse}).
\label{tab:metal}}
\begin{tabular}[h]{ccccc}
\noalign{\smallskip} \hline \hline \noalign{\smallskip}
$X_{0}$&$Y_{0}$&$Z_{0}$&[Fe/H]&$\alpha_{\rm ov}$\\
\noalign{\smallskip} \hline \noalign{\smallskip}

0.735&0.250&0.015&$-$0.0794\,\,\,\,&0.0--0.2\\

0.685&0.300&0.015&$-$0.0488\,\,\,\,&0.0--0.2\\

0.700&0.280&0.020&0.0668&0.0--0.2\\

0.725&0.250&0.025&0.1484&0.0--0.2\\

0.675&0.300&0.025&0.1794&0.0--0.2\\

\noalign{\smallskip} \hline  \noalign{\smallskip}
\end{tabular}
\end{center}
\end{table}

Although the final comparison with observations is done
considering rotating models, our procedure requires the
computation of non-rotating models and the corresponding
isochrones. Hence, sequences of non-rotating models were
calculated for masses between 0.8 $M_{\odot}$ and 5.0 $M_{\odot}$,
from the ZAMS to the sub-giant branch and the corresponding
isochrones computed for each sets of parameters in
Table~\ref{tab:metal}. In the following analysis we shall consider
three ages, $A$, of 70 Myr, 100 Myr and 130 Myr. Finally, for
comparison with the observations three distance moduli, $d$, of
5.50, 5.60 and 5.70 mag were considered
 except for models with [Fe/H]=$-$0.0488, in which case 
an additional value of $d=5.39$ mag was used.

Figure~\ref{fig:dscutis} shows an example of such isochrones
computed with the following parameters: [Fe/H]=0.0668,
$\alpha_{\rm ov} = 0.2$, $A = 70$ Myr and $d= 5.70$ mag. The
isochrones were calibrated from [$T_{\rm eff}, \log
(L/L_{\odot})$] to $(B-V, M_{V})$ by using the Schmidt--Kaler
(\cite{schmidt}) calibration for magnitudes and the relationship
between $T_{\rm eff}$ and $B-V$ of Sekiguchi \& Fukugita
(\cite{sekiguchi}) for the colours. As can be seen in
Fig.~\ref{fig:dscutis}, in the case illustrated here the isochrone
matches the observed colour--magnitude diagram.  For some combinations 
of high metallicity and low
distance modulus the isochrone fit is not satisfactory but we have
not rejected those combinations to allow for an independent
determination of the cluster parameters from the oscillation
frequencies.   
Hence we have considered 
the entire combination of cluster parameters given in
Table~\ref{tab:metal} and the three ages given above.

\subsection{Photometric effect of rotation on the HR diagram}

It is known that fast rotation affects the position of a star in the
colour--magnitude diagram (e.g.\ Tassoul \cite{tassoul}).
In particular a ZAMS rotating model has a smaller luminosity and mean
$T_{\mathrm{eff}}$ than a non-rotating model with the same mass and chemical
composition.
Also, the magnitude and the colour of a rotating star depend on
the aspect angle, $i$, between the line of sight and the rotation
axis. In particular, a rotating star seen pole-on
appears brighter and hotter than the same star seen equator-on.

We now search for the range of masses and angular rotations suitable for
each star. To do this we shall use the
results of P\'erez Hern\'andez et al.\ (\cite{perez2}), in which the
correction to the photometric magnitudes of non-rotating models needed to obtain
those of rotating models were computed for main sequence stars
of spectral type A0 to F5.
In this calculation, we shall take into account the observed $v \sin i$ for each star.
As an example,
Fig.~\ref{fig:cor} illustrates the corrections due to rotation in
the particular case of the $\delta$ Scuti type star V650 Tau with $v \sin
i=175$ km s$^{-1}$. The actual position of the star is shown with
a thick dot upon which there is a cross that gives the errors in
the estimation of $(M_{V})_{0}$ and $(B-V)_{0}$ calculated
according the following expressions:

\begin{figure}
\resizebox{\hsize}{!} {\includegraphics[]{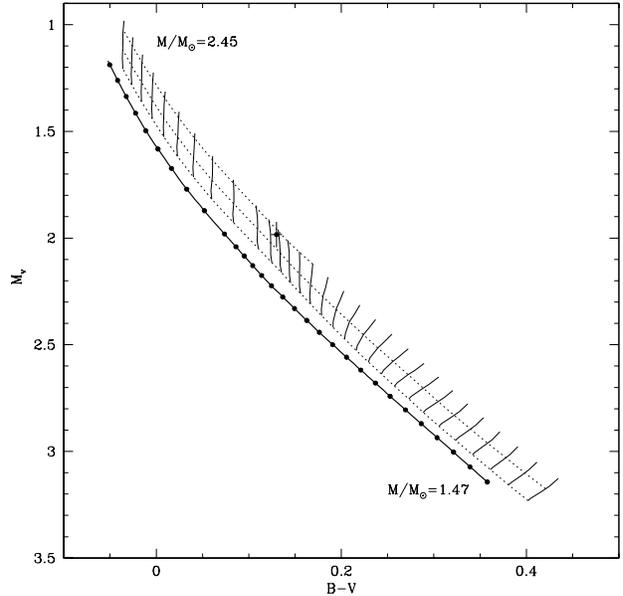}}
\caption{Colour--magnitude diagram illustrating the photometric
corrections due to rotation in the case of V650 Tau
($v\sin i = 175\,$km s$^{-1}$). The  thick continuous line is the
same isochrone represented in Fig.~\ref{fig:dscutis} ([Fe/H] =
0.0668, $\alpha_{\rm ov} = 0.2$, $A = 70$ Myr, $d=5.7$ mag).
Some models are shown with filled circles upon the isochrone. The masses of the 
first and last models are indicated.
The small tracks represent the photometric corrections due to rotation for each  
 the models indicated on the isochrone when $i$ runs from
$i=90^{\circ}$ to $i=i_{\rm min}$. The corrections at fixed $i$ are represented by dotted lines.}
\label{fig:cor}
\end{figure}

\begin{equation}
\sigma[(B-V)_{0}] = \sqrt{\sigma^{2}(B-V) +\sigma^{2}[E(B-V)]},
\end{equation}

\begin{equation}
\sigma[(M_{V})_{0}] = \sqrt{\sigma^{2}(d)+\sigma^{2}(m_{V}) +
(3.2\sigma[E(B-V)])^{2}}
\, ,
\end{equation}

\noindent
where $\sigma(d)=0.06$ is the error in the distance modulus. The
errors in the magnitude $M_{V}$, colour index $(B-V)$ and reddening
$E(B-V)$ are given in Table~\ref{tab:stars}.

The range of masses and rotational velocities of the star depends
on the stellar parameters considered and hence needs to be computed
for each combination in Table~\ref{tab:metal}. In particular, in
the example illustrated in Fig.~\ref{fig:cor}, the same cluster
parameters as in Fig.~\ref{fig:dscutis} were considered
([Fe/H]=0.0668, $\alpha_{\rm ov} = 0.2$, $A = 70$ Myr and $d=
5.70$ mag). The isochrone is represented by a thick continuous
line and some of its models are shown by dots. For each model on
the isochrone there is a small track corresponding to the
photometric corrections due to rotation computed by using the
projected velocity  $v \sin i$ of the star and changing the angle
of inclination from $i=90^{\circ}$ to the minimum angle $i=i_{\rm
min}$ corresponding to $\sim 0.90 \Omega_{\rm br}$, where
$\Omega_{\rm br}$ is the break-up angular rotational velocity.
This angular velocity is given by the balance between the
gravitational attraction and the centrifugal force at the equator.
The dotted lines give the corrections at fixed angles:
$i=90^{\circ}$ (the closest line to the isochrone), $i=50^{\circ}$
(intermediate line) and $i=39^{\circ}$ (the reddest line). It is
evident that the actual position of the star, including the error
box, is associated to a domain of non-rotating models. In this
case we have obtained for the  star considered a range
of masses of $[1.86,1.91]M_{\odot}$ and a range of angles of
inclination  of $[39^{\circ},50^{\circ}]$.  Nonetheless, in order
to take into account possible errors coming from the isochrone
calibration we shall consider a wider range of aspect angles
$i=[i_{\rm min}, 90^{\circ}]$. The same procedure is carried out
for the other stars.  Also one needs to consider all the
metallicities, ages, overshooting parameters and distance moduli for
the whole set of stars.

Having obtained ranges of $M$ and values of $i_{\rm min}$, we proceed to find the
corresponding range of angular rotational velocities $\Omega$, for
which an estimate of the equatorial radius, $R_{\rm eq}$, of the
rotating model is needed. Under the assumption of uniform rotation
and approximating the surface of the star by a Roche surface (see
e.g. P\'erez Hern\'andez et al. \cite{perez2}) one has:

\begin{equation}
R_{\rm eq} \simeq R_0 \frac{3}{\omega} \cos \left\{ \frac{1}{3}
\left[ \pi + {\rm arcos}\, \omega \right] \right\}
\; ,
\end{equation}

\noindent where $\omega \simeq \Omega/\Omega_{\rm br}$,
$\Omega^2_{\rm bq} \simeq 8GM/(27R^3_0)$ and $R_0$ is the polar
radius. As noted in P\'erez Hern\'andez et al.~(\cite{perez2}), the
polar radius can be approximated by the radius of a spherical
non-rotating model with the same mass and evolutionary state
(since our stars are close to the ZAMS, a non-rotating model of
the same mass and age is suitable for this error-box estimation).

Since on the other hand the angular rotation is related to the equatorial radius by

\begin{equation}
\Omega \simeq \frac{(v\sin i)_{\rm obs}}{R_{\rm eq}\sin i}
\; ,
\end{equation}

\noindent it is possible to carry out a simple iterative process
to obtain a range of possible rotations, $[\Omega_{\rm min},\Omega_{\rm max}]$, 
from the previously obtained range of
angle of inclinations $[i_{\rm min},90^{\circ}]$ and masses.
Since for given data parameters, the mass of the star is obtained with an 
uncertainty of $\sim 0.02$ $M_{\odot}$ this
iterative process was carried out for just one mass.

\begin{figure}[!t]
\resizebox{\hsize}{!} {\includegraphics[width=22cm]{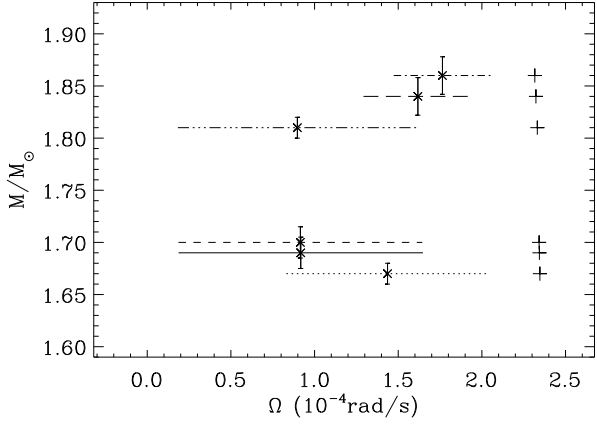}}
\caption{Range of rotation rates against masses estimated for the
six stars from the photometric corrections due to rotation applied
to the same isochrone depicted in Figure~\ref{fig:cor}. The vertical error
bars give the estimated range of mass for each star. The crosses
give the values of the break-up angular rotational velocity,
$\Omega_{\rm br}$. The ranges of $\Omega$ and $M$ are associated
with stars as follows: dot-dashed line for V650 Tau $(v\sin i=175$
km s$^{-1})$; long-dashed line for HD 23628 $(v\sin i=150\,$km
s$^{-1})$; three-dots dashed line for HD 23194 $(v\sin i=20$
km s$^{-1})$; dashed-line for V647 Tau $(v\sin i=20$ km s$^{-1})$;
continuous line for V624 Tau $(v\sin i=20\,$km s$^{-1})$ and
dotted line for V534Tau $(v\sin i=90\,$km s$^{-1})$. }
\label{fig:om}
\end{figure}

As an example, Figure~\ref{fig:om} shows the ranges of $\Omega$
obtained for the six stars as a function of the mass of the models
considering the same cluster parameters as in
Figure~\ref{fig:cor}. Lines of different types give the estimates
for the different stars as indicated in the figure. A similar
procedure needs to be considered for all sets of parameters in
Table~\ref{tab:metal}.

With the information obtained above we compute evolutionary
sequences of rotating models with the same input physics as the
non-rotating models but with appropriate initial angular
rotational velocities in order to match as closely as possible the
extreme values in the interval $[\Omega_{\rm min},\Omega_{\rm
max}]$ at the final age. Solid-body rotation and conservation of
global angular momentum during the evolution were assumed during
the calculus. In order to have a better overview of the rotation rates of
the stars we have also computed models with an intermediate value
of $\Omega_{\rm mid} \approx 0.5 \Omega_{\rm br}$. A total amount
of 1620 sequences of rotating models were finally obtained for the
whole ensemble of stars.

\subsection{Theoretical oscillation frequencies}

The eigenfrequencies of the rotating models previously described
have been calculated using the oscillation code FILOU (Tran Minh
et al.\ \cite{tran}; Su\'arez  \cite{suarez2}). We have considered
frequency perturbations up to second order in the rotation rate.
Frequencies of the modes are labelled in the usual way: $n, l, m$
for the radial order, degree and azimuthal order. The radial
orders of a given mode are assigned according to the Scuflaire
criterion (\cite{scuflaire}) with $n>0$ for the $p$ modes, $n<0$
for the $g$ modes, $n=1$ for the fundamental radial mode of degree
0 and 1, and $n=0$ for the fundamental mode with the $l=2$.
Eigenfrequencies were computed up to $l=2$. For geometrical
reasons higher degree modes are expected to have considerably
smaller amplitudes. The theoretical frequencies cover the
frequency range of the observed pulsation peaks (see
Table~\ref{tab:freq}). Coupling between modes is not considered in
the present work.

The estimated interval of $\Omega_{\rm
rot}$ for all the stars may be as large as 
$\Delta \Omega_{\rm rot} \sim 1.6 \times 10^{-4}$ rad/s 
which in terms of cyclic rotational frequency corresponds
to $\Delta \nu_{\rm rot} \sim  25\,\mu$Hz (see for instance
Fig.~\ref{fig:om}). After some tests, we found that a
satisfactory comparison between the observed and theoretical
frequencies cannot be achieved by using only the frequencies
computed for models with three $\Omega_{\rm rot}$ within this
interval. To overcome this difficulty once we had the mode
frequencies for each series of models (three $\Omega_{\rm rot}$
for fixed $M$, [Fe/H], $\alpha_{\rm MLT}$, $\alpha_{\rm ov}$, $d$
and $A$), we proceeded to interpolate the results on $\Omega_{\rm
rot}$ in 21 equally spaced points covering the interval
[$\Omega_{\rm min}, \Omega_{\rm max}$]. A quadratic spline
interpolation was applied. In order to
evaluate the goodness of the interpolation we also computed the
eigenfrequencies for several randomly selected models. We have
found that there is a good agreement between the interpolated  and
theoretical frequencies up to $\nu \sim 600\, \mu$Hz ($n \sim 6$),
 the average of absolute separation being $|\nu_{\rm int} -\nu_{\rm
cal} |\sim 0.3\,\mu$Hz. The disagreement found beyond this value can be
explained by the fact that the second order effect of rotation is
enhanced at higher frequencies. In any case, as can be seen in
Table~\ref{tab:freq}, the highest frequency we are dealing with
is $\simeq 574\,\mu$Hz.

\section{Method of comparison \label{sec:comp}}

Once we have the mode frequencies for all sets of parameters we
compare the observational frequencies ($\nu_{\rm obs}$)
with the theoretical ones ($\nu_{\rm cal}$) at each interpolated
$\Omega_{\rm rot}$.

Let us consider a rotating model with given parameters for just
one star. Then, for all the possible combinations between the
observed and computed frequencies, we compute the quantity $\chi^{2}$ given by

\begin{equation}
\chi^{2}= \frac{1}{N}\sum^{N}_{j=1}(\nu_{{\rm obs},j} - 
\nu_{{\rm cal},j})^{2}
\, ,
\end{equation}

\noindent
where $N$ is the number of observational frequencies.
In this computation it is understood that each theoretical frequency is
assigned to one observed frequency at most.

\begin{figure}[!ht]
\resizebox{\hsize}{!} {\includegraphics[width=22cm]{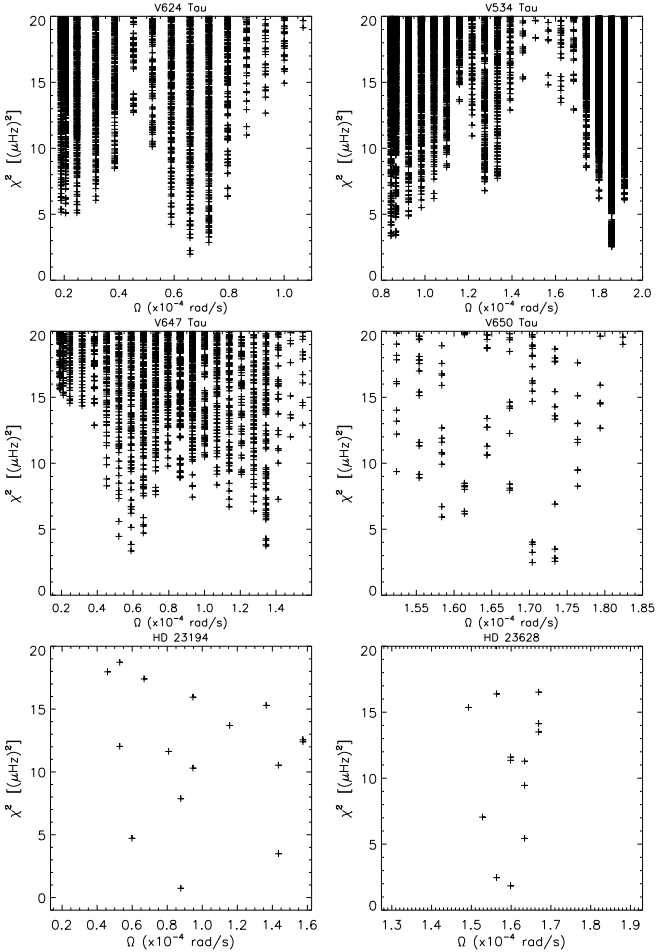}}
\caption{$\chi^{2}$ against $\Omega$ for six models with common parameters
 [Fe/H], $\alpha_{\rm ov}$, $d$ and $A$.}
\label{fig:ang}
\end{figure}

As a first step, in order fully to  exploit the collective behaviour
of stars within open cluster we consider the solutions of the six
stars with given cluster parameters of [Fe/H], $\alpha_{\rm ov}$,
$d$ and $A$. The models corresponding to each star differ in the
mass and the rotation rate. As an example,  Fig.\ref{fig:ang}
shows $\chi^{2}$ against  angular rotational velocity,
$\Omega$, for the six models corresponding to the same parameters
as in Fig.~\ref{fig:cor} but without overshooting ([Fe/H]=0.0668,
$\alpha_{\rm ov}=0.0$, $d$=5.70 mag and $A$=70 Myr). Each panel
corresponds to the star indicated in the figure. For clarity,
only solutions with $\chi^{2}< 20 \, (\mu\rm{Hz}^{2})$ are shown. Similar
features are found for other metallicity, distance moduli and ages,
but the lowest $\chi^{2}$ may appear with rather high values.

Since we expect the same [Fe/H], $d$ and $A$ for all the stars in
the cluster we shall assume that the best fits should happen
simultaneously in the six stars despite
 differences in rotation
rate and masses of the models. Hence we proceed to calculate the
mean square root of the lowest $\chi^{2}$ found in each model for
given common parameters by means of the following expression:

\begin{equation}
\epsilon= \sqrt{\frac{1}{6}\sum^{6}_{i=1}(\chi_{{\rm min},i})^{2}}.
\end{equation}

\begin{figure}
\resizebox{\hsize}{!} {\includegraphics[width=22cm]{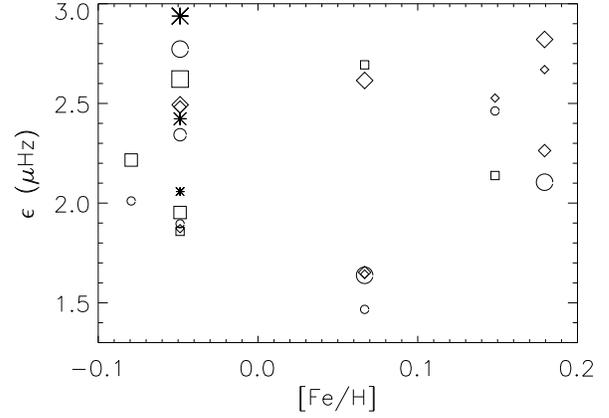}}
\caption{Minimum values of the mean square root of the difference
between observed and theoretical frequencies for models with the
same values of [Fe/H], $\alpha_{\rm ov}$, $d$ and $A$  against
[Fe/H]. For clarity, only models without overshooting are shown.
The symbols are related to different values of $A$ and $d$ as
follows:  asterisks ($d=5.39$ mag), squares ($d=5.50$ mag), diamonds ($d=5.60$ mag), circles
($d=5.70$ mag), small (70 Myr), middle (100 Myr) and big (130
Myr). } \label{fig:sig}
\end{figure}

\noindent Figure~\ref{fig:sig} shows $\epsilon$ against
[Fe/H] for models without overshooting.  For the sake of clarity
only the points with $\epsilon \leq 3.0\,\mu$Hz are shown. It
can be seen that the solution with the smallest $\epsilon $
corresponds to [Fe/H]=0.0668. A plot of similar characteristics is
obtained for models with overshooting.

Since only those identifications obtained from eq.~(5) with low
$\chi^{2}$ values are of interest, we require that for each star
the solution must have $\chi \leq 3.5\,\mu$Hz. 
In Fig.~\ref{fig:sig} we reject those solutions that have at least one
model with $\chi> 3.5\,\mu$Hz for all the parameters and
combinations between observed and theoretical frequencies. With
this restriction we found that only those solutions associated with
the models represented by the four lowest points in
Fig.~\ref{fig:sig} should be considered.  Similar
solutions were found for models with overshooting. All these models are listed
in Table~\ref{tab:dom} and  will be analysed in detail later.

\begin{table}[]
\begin{center}
\caption{Ensembles of models with and without overshooting, which,
after applying a threshold of $\chi \leq 3.5\,\mu$Hz, have potential valid
solution (see text for details).}
\begin{tabular}{cccccc}
\noalign{\smallskip}
\hline
\noalign{\smallskip}
\multispan6 \hfill [Fe/H] = 0.0668 \hfill \\
\noalign{\smallskip}
\hline
\noalign{\smallskip}
\multispan3 \hfill $\alpha_{\rm ov} = 0.00$ \hfill &\multispan3 \hfill
$\alpha_{\rm ov} = 0.20$ \hfill \\\noalign{\smallskip}
\hline
\noalign{\smallskip}
& $d$ & $A$ &   &  $d$ &   $A$ \\

& (mag)&(Myr)& & (mag)&(Myr)          \\
\noalign{\smallskip}
\hline
\noalign{\smallskip}
C1&  5.70 & \,\,70&D1& 5.70 & \,\,70 \\
C2& 5.70 & 130& D2&5.60 & 100\\
C3&     5.60 & \,\,70 &D3&5.60 &130 \\
C4&     5.60 & 100&D4&5.70 &100 \\

\noalign{\smallskip}
\hline
\noalign{\smallskip}
\end{tabular}
\normalsize
\end{center}
\label{tab:dom}
\end{table}

\begin{figure}
\resizebox{\hsize}{!} {\includegraphics[width=22cm]{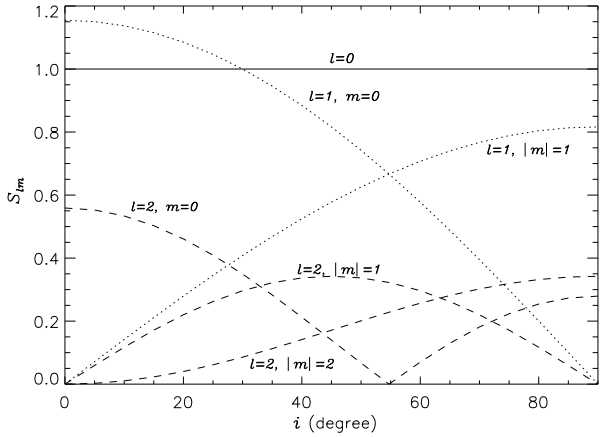}}
\caption{Variation of visibility amplitude with the inclination angle for radial
$l=0$ (continuous line) and non-radial $l=1$ (dotted lines) and $l=2$ (dashed lines)
modes.}\label{fig:vis}
\end{figure}

We shall use a geometrical argument  to  reduce further the
number of solutions so far available for each star. To this end,
we take into account that the visibility of each mode
depends on the angle of inclination. 
Following Pesnell~(\cite{pesnell}), we illustrate in
Fig.~\ref{fig:vis} the spatial response function
$S_{lm}$ against $i$, for mode degree $l=0,1,2$. For
simplicity, limb-darkening has been neglected.  It can be seen that for $i \approx
90^{\circ}$ only even $l+m$ values can be detected, while for $i
\approx 0^{\circ}$ only modes with $m = 0$ will be visible.

We introduce an hypothesis based on the visibility of the modes:
we  reject any solution with one or more modes
with $S_{lm}^{\prime} < 0.1$. After applying this constraint we
were able to limit the analysis to four sets in 
Table~\ref{tab:dom}, two with $\alpha_{\rm ov}=0.0$: C1
$(d=5.70,\, A=70)$, C3 $(d=5.60,\, A=70)$ and two with
$\alpha_{\rm ov}=0.2$: D1 $(d=5.70,\, A=70)$, D2 $(d=5.60,\, A=100
)$. We note that while for the stars HD 23628 and HD 23194 only
a few identifications ($<$ 8) remain, the number of identifications
for V624 Tau, V534 Tau, V647 Tau and V650 Tau remained larger than
$100$ in most  cases.

\section{Results and discussion \label{sec:results}}

In order to discuss the results we shall introduce a parameter
 $\Delta$ associated with each star in each identification and given
by

\begin{equation}
\Delta = {\rm max} ( |\nu_{\rm obs}- \nu_{\rm cal}| ).
\end{equation}

Table~\ref{tab:delta} lists the number of solutions for the six
stars in each ensemble. Different maximum values of $\Delta$ have
been considered. Certain features can be derived from these
general results:

\begin{table*}[]
\begin{center}
\caption{Number of possible solutions for models in each ensemble
for different values of $\Delta$ (in $\mu$Hz). \label{tab:delta}}
\begin{tabular}{lcccccccccc}
\noalign{\smallskip}
\hline
\noalign{\smallskip}
&\multispan{10} \hfill [Fe/H] = 0.0668 \hfill \\
\noalign{\smallskip}
\hline
\noalign{\smallskip}
&\multispan5 \hfill $\alpha_{\rm ov} = 0.00$ \hfill &\multispan5 \hfill $\alpha_{\rm ov} = 0.20$ \hfill \\\noalign{\smallskip}
\hline
\noalign{\smallskip}
&$\Delta\,(\mu{\rm Hz})\, \leq$  &1.0 & 2.0&3.0 &3.5 &  $\Delta\,(\mu{\rm Hz})\, \leq$    &1.0   &2.0 & 3.0 & 3.5   \\\noalign{\smallskip}
\hline
\noalign{\smallskip}
&$\frac{M}{M_{\odot}}$&  & & &&$\frac{M}{M_{\odot}}$&  & & & \\
\noalign{\smallskip}
\hline
\noalign{\smallskip}
&C1  & & & & &D1&  & & & \\
V624 Tau&1.70 &0 &0 &12 &18&1.69 &0 &0 &0 &12 \\
V534 Tau&1.67 &0 &0 &0 &16&1.67 & 0& 0& 0&0 \\
V647 Tau&1.70 &0 &0 &0 &2 &1.70 &0 & 0&0 &2 \\
V650 Tau&1.86 &0 &0 &7 &13&1.86 &0 &2 &4 &4 \\
HD 23194&1.80 &0 &1 &2 & 3&1.81 &0 &0 &3 &3 \\
HD 23668&1.84 &0 &1 &2 & 4&1.84 &0 &0 &2 &3 \\
\noalign{\smallskip}
\hline
\noalign{\smallskip}
&C3  & & & & &D2 & & & & \\
V624 Tau&1.68 &0 &0 &6 &12&1.68 &0 &0 &0 &0\\
V534 Tau&1.66 &0 &0 &0 &24&1.66 & 0& 0& 0&0 \\
V647 Tau&1.69 &0 &0 &0 &2 &1.69 &0 & 0&0 &3 \\
V650 Tau&1.86 &0 &0 &0 &0&1.86 &0 &0 &0 &3\\
HD 23194&1.79 &1 &1 &2 &2&1.79 &1 &1 &1 &2 \\
HD 23668&1.82 &0 &0 &3 &4&1.84 &0 &2 &3 &6 \\
\noalign{\smallskip}
\hline
\noalign{\smallskip}
\end{tabular}
\normalsize
\end{center}
\end{table*}

\begin{itemize}
\item The resulting masses for each star are similar
for all the solutions. 

\item The number of solutions for a given star at fixed $\Delta$
associated with models with overshooting is smaller than without it.

\item There are no solutions with $\Delta \leq 3.0\,\mu$Hz for the stars V534 Tau
and V647 Tau, while the star HD 23194 shows solutions even with
$\Delta \leq 1.0\mu$Hz.

\item In order to find a set of models that has at least one solution simultaneously
for all the stars a value of $\Delta \simeq 3.5\,\mu$Hz is needed.
This solution is found for C1 that corresponds to the cluster
parameters [Fe/H]=0.0668, $\alpha_{\rm ov}=0.0$, $d=5.70$ and
$A=70$, and it coincides with the minimum in Figure~\ref{fig:sig}.

\item
The remaining ensembles in Table~\ref{tab:delta} have at least
one solution simultaneously when
the value of $\Delta$ is increased slightly. In particular, for
C3 a value of $\Delta \simeq 4.5\,\mu$Hz is needed to obtain a
solution while for D1 and D2 a value of $\Delta \simeq 3.8\,\mu$Hz
is required. 

\end{itemize}

We have found that the
identifications and the range of aspect angles derived for each star are 
similar for all the solutions. 
This agreement may be due to the fact that the estimated range of mass and radius,
and hence of mean densities, is similar for all the solutions. 
At the  ages considered the overshooting has a negligible
effect on the models.

Table~\ref{tab:results} summarizes the parameters estimated for
the cluster as well as for each star associated with the possible
identifications. The radial orders of these
modes are consistent with growth rate predictions
(Su\'arez et al.\ \cite{suarez}). 
While these identifications cannot be considered as
definitive but rather as ``best fit solutions'', some
information on the stars could be derived.
On the one hand, those stars with  smaller masses
(V624 Tau, V534 Tau and V647 Tau) have more than six frequencies and could have 
simultaneously excited radial and non-radial oscillations although
at most two radial modes. 
Following Hern\'andez et al.\ (\cite{hernandez}), 
at most two radial modes were found to be excited for the $\delta$
Scuti stars in the Praesepe cluster.
On the other hand, the most massive stars, V650 Tau, HD
23628 and HD 23194, with a mean mass of 1.83 $M_{\odot}$
present on average few observational frequencies ($N \le 4$) that
might be  associated only with non-radial modes. With the
present observational data it remains unclear, however, whether this
is a general trend for $\delta$ Scuti stars. 

\begin{figure}
\resizebox{\hsize}{!} {\includegraphics[width=22cm]{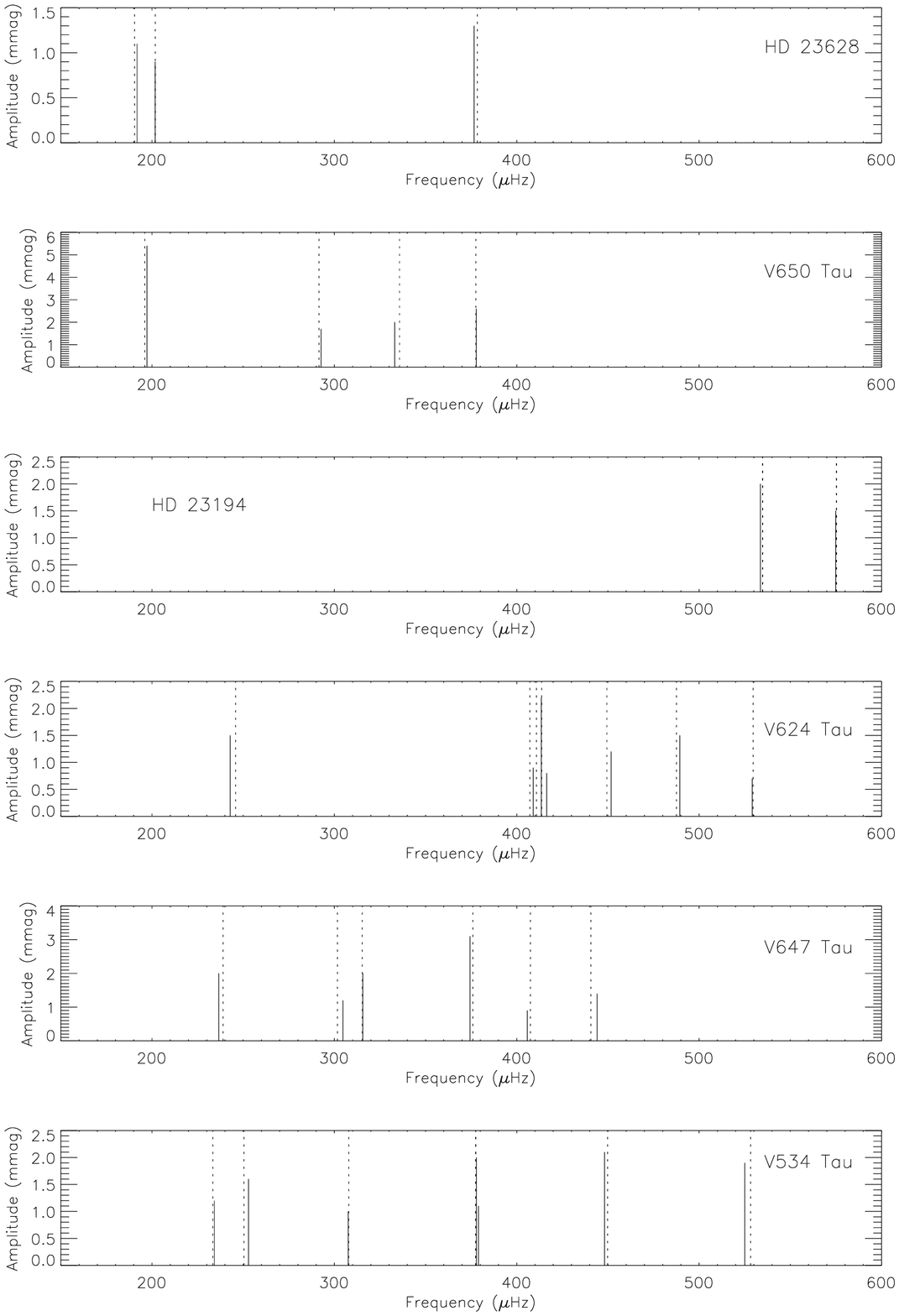}}
\caption{Comparison of observed (solid lines) and theoretical (dot lines)
frequencies of the best fit solutions in C1-models (Table~\ref{tab:delta}).}
\label{fig:comp}
\end{figure}

\begin{center}
\begin{table*}[]
\caption{Summary of parameters derived for the Pleiades cluster as well as for
$\delta$ Scuti stars with the possible identification of the frequency modes.
\label{tab:results}}
\begin{tabular}{lcclcc}

\noalign{\smallskip}
\hline
\hline
\noalign{\smallskip}
 \multispan{6}\hfill [Fe/H] = 0.0668, $\alpha_{\rm ov} = $[0.00-0.20], $m_{V}-M_{V}=$[5.60-5.70], $A=$[70-100]${\times}10^{6}$ years \hfill \\
\noalign{\smallskip}
\hline
\hline
\noalign{\smallskip}
 Star&$\nu $ &Identification &Star&$\nu $ &identification  \\
         &($\mu$Hz)& ($n,l,m)$&        &($\mu$Hz)& ($n,l,m)$  \\
\noalign{\smallskip}
\hline
\hline
\noalign{\smallskip}
V624 Tau&242.9& (1,0,0) &V650 Tau& 197.2&(1,1,1) \\
$M = [1.68$-1.72]\,$M_{\odot}$         &409.0& (3,1,1)    &$M =$ [1.84-1.88]\,$M_{\odot}$       & 292.7&(0,2,$-$2),(2,2,2)   \\
  $\nu_{\rm rot}= [3$-5]\,$\mu$Hz &413.5& (3,1,0) & $\nu_{\rm rot}= [$25-28]\, $\mu$Hz  & 333.1&(3,1,1),(3,2,2)  \\
 $i=[37^{\circ}$-$67^{\circ}]$&416.4&(3,1,$-$1) & $i=[60^{\circ}$-$70^{\circ}]$ & 377.8&(2,2,$-$2),(3,1,0)  \\
& 451.7& (3,2,$-$2),(4,0,0) & &&(3,1,$-$1),(3,2,1)\\
                &489.4 & (4,1,0),(4,1,1) &&&\\
              &529.1 & (4,2,$-$1),(4,2,$-$2)&&&\\
            & &(5,0,0) &&&\\
\noalign{\smallskip}
\hline
\noalign{\smallskip}
V534 Tau&234.2& (1,1,1)&HD 23628 &191.8 &(0,2,2)   \\
$M = [1.65$-1.69]\,$M_{\odot}$        &252.9& (1,1,0)&$M = [1.82$-1.86]\,$M_{\odot}$ &201.7 & (1,1,1)    \\
$\nu_{\rm rot}= [14$-16]\,$\mu$Hz &307.6&(2,0,0),(2,1,1)& $\nu_{\rm rot}= [24$-26]\,$\mu$Hz  &376.6&(2,2,$-$2)\\
$i=[59^{\circ}$-$79^{\circ}]$  &377.9& (2,2$-$1),(3,0,0)   &$i=[53^{\circ}$-$59^{\circ}]$&&\\
         &379.0 & (2,2$-$1),(3,0,0)  &  &&\\
         & & (3,1,1)  &&\\
        &448.1& (3,2,$-$1),(4,0,0)      &&&\\
               &525.0& (4,2,$-$1)  &&&\\
\noalign{\smallskip}
\hline
\noalign{\smallskip}
V647 Tau &236.6&(1,1,1)  &HD 23194&533.6&(5,1,1),(5,1,$-$1) \\
 $M = [1.68$-1.72]\,$M_{\odot}$  &304.7&(1,2,1) &$M = [1.78$-1.82]\,$M_{\odot}$&574.9&(5,2,0),(5,2,1)  \\
 $\nu_{\rm rot}= 10-11\,\mu$Hz&315.6&(2,0,0),(2,1,1)& $\nu_{\rm rot}= [14$-23]\,$\mu$Hz & & \\
 $i=17^{\circ}$-18$^{\circ}$ &374.4&(2,2,$-$1)&$i=7^{\circ}$-$12^{\circ}$ &&\\
     &405.8&(3,1,0)  &&&\\
         &444.1&(3,2,0)    &&&\\
\noalign{\smallskip}
\hline
\hline
\noalign{\smallskip}
\end{tabular}
\end{table*}
\end{center}

A comparison of observed and computed frequencies is shown in
Figure~\ref{fig:comp}. In each panel the solid lines represent the 
observational frequencies with their corresponding amplitudes.
The ``best fit'' theoretical frequencies
are represented by dotted lines. The stars are arranged from top to bottom
according to their magnitude. 

From this figure it follows that
there is an asymmetric triplet in the observational spectrum of
V624 Tau. We have found that for all the solutions the
identification is always a $(l=1,\,n=3)$ mode split by rotation.
In turn, this implies that an independent estimate of the mean rotation
rate can be obtained from the rotational splitting, up to second
order, given by

\begin{equation}
\frac{\nu_{-m} - \nu_{m}}{2 m} \sim \frac{\Omega_{\rm rot}}{2 \pi}
(1-C_{nl}),\;\;\;\; m=1,2,\ldots,l
\end{equation}

\noindent
We have found complete agreement between the cyclic
rotational frequency computed from the above expression and those
derived from the mode identification,  the differences being 
$\sim 0.03-0.04\,\mu$Hz. We have also found a fairly
good agreement between the observations and the models in the differences
$\nu_{-m}-\nu_{0}$ and $\nu_{+m}-\nu_{0}$.

\medskip
If the estimated inclination angles for
the stars are correct, all of them could be rapid rotators ($\nu_{\rm rot}>10\mu$Hz)
except for V624 Tau with $\nu_{\rm rot}\leq 5\mu$Hz.
HD 23194 and V647 Tau, whose projected rotational velocities are low,
$v\,\sin\,i = 20$ kms$^{-1}$, would have equatorial velocities as high as
90 and 70 kms$^{-1}$ respectively.

\medskip
In the present work the distance is found to be
$m_{V}-M_{V} = 5.60-5.70$ mag $(132 - 138$ pc) and agrees very well
with that derived recently
by independent determinations using independent techniques and data
(Pan et al.\ \cite{pan}; Munari et al.\ \cite{munari}; Soderblom et al.\
\cite{soderblom}; Percival et al.\ \cite{percival}).
 However, although
the isochrones computed with sub-solar metallicity
  ($Z=0.015$ and $Y=0.30$) match well the observational data
 with a distance modulus of $d=5.39$ mag in agreement with
that of HIPPARCOS within 1-$\sigma$ error,
the resulting solutions (shown by asterisks in
Fig.~\ref{fig:sig}) do not lead to
 good fits.  Moreover, the plot of $\epsilon$ against $d$ which we do
not reproduce here, shows better fits for further distances.
For solar metallicity, in order to reproduce the HIPPARCOS MS of the Pleiades,
unrealistically high helium abundances are required (e.g. $Y=0.34$ Belikov et al.
\cite{belikov}, $Y=0.37$ Pinsonneault et
 al. \cite{pinsonneault}).

\section{Conclusions \label{sec:conclusions}}

In this study we have performed a seismological
analysis of six $\delta$ Scuti stars belonging to the Pleiades
cluster to identify their frequency modes.
To the best of our knowledge this group of
variables constitutes the most statistically significant sample of
$\delta$ Scuti stars analysed in an open cluster to date.

Rotational effects  were considered in different stages of
the analysis: first when determining the star positions in the HR
diagram and second when computing stellar models that approximately
reproduce the evolutionary stage of the stars, and finally when
computing theoretical oscillation frequencies in order to
construct seismic models for target stars. The comparison between
observational and computed frequencies was carried out by
a least-squares fit.
There is a large number of possible solutions partly because
 neither the equatorial velocity nor the inclination
angle are known a priori. In order to limit the number of possible
solutions we used the relationship between the
amplitude visibility, $S_{lm}$ and the aspect angle, $i$.
As a result we found  few solutions for each star,
suggesting the existence of only $p$ modes of low degree, low radial order
in all the stars. For the less massive stars,
solutions with at most two radial modes were also possible. These solutions have
associated ranges of stellar parameters for each star. Most of the
stars could be rapid rotators according to the estimated angle of inclination,
$i$.

The best fits between observational and theoretical frequencies
are achieved for global cluster parameters of [Fe/H] = 0.0668
$(Z_{0}=0.02,\, X_{0}=0.70)$, $A=70 - 100\,$Myr and $d=5.60 -
5.70$ mag. The derived distance modulus for the Pleiades cluster
agrees with that of the main sequence fitting method, in spite of the fact that
the isochrones with sub-solar metallicity
closely matches the Pleiades HR diagram with the HIPPARCOS distance.

\begin{acknowledgements}
This work has been partially funded by grants AYA2001-1571, ESP2001-4529-PE and 
ESP2004-03855-C03-03 of the Spanish national research plan. L.F.M acknowledges the partial 
financial support by the Spanish AECI. J.C.S. acknowledges the financial support by the Spanish Plan of Astronomy and Astrophysics, under project AYA2003-04651, by the Spanish ``Consejer\'{\i}a de Innovaci\'on, Ciencia y Empresa'' from the ``Junta de Andaluc\'{\i}a local government, and by the Spanish Plan Nacional del Espacio under project ESP2004-03855-C03-01.
\end{acknowledgements}

\end{document}